\documentclass[12pt]{article}
\usepackage{epsfig}

\newlength{\abstwidth}
\newenvironment{Itemize}{\begin{list}{$\bullet$}%
{\setlength{\topsep}{0.2mm}\setlength{\partopsep}{0.2mm}%
\setlength{\itemsep}{0.2mm}\setlength{\parsep}{0.2mm}}}%
{\end{list}}
\newcounter{enumct}
\newenvironment{Enumerate}{\begin{list}{\arabic{enumct}.}%
{\usecounter{enumct}\setlength{\topsep}{0.2mm}%
\setlength{\partopsep}{0.2mm}\setlength{\itemsep}{0.2mm}%
\setlength{\parsep}{0.2mm}}}{\end{list}}

\begin{document}

\sloppy

\begin{flushright}
LU TP 00--54\\
hep-ph/0012188\\
December 2000
\end{flushright}

\vspace{2cm}

\begin{center}
{\LARGE\bf PYTHIA Status Report%
\footnote{to appear in the Proceedings of the Linear Collider %
Workshop 2000, Fermilab, October 24--28, 2000}}\\[10mm]
{\Large T. Sj\"ostrand\footnote{torbjorn@thep.lu.se}} \\[3mm]
{\it Department of Theoretical Physics,}\\[1mm]
{\it Lund University,}\\[1mm]
{\it S\"olvegatan 14A,}\\[1mm]
{\it S-223 62 Lund, Sweden}
\end{center}
 
\vspace{2cm}
 
\begin{center}
{\bf Abstract}\\[2ex]
\begin{minipage}{\abstwidth}
Recent improvements in the \textsc{Pythia} event generator are
summarized: new hard subprocesses, $\gamma^*\gamma^*$ physics, 
QCD final-state showers, and more.
\end{minipage}
\end{center}
 
\vspace{1cm}

\noindent 
\rule{160mm}{0.5mm}

\vspace{1cm}

Since 1997, the \textsc{Jetset}~7.4, \textsc{Pythia}~5.7 
\cite{pythiaold} and  \textsc{SPythia} \cite{spythia} programs 
have been joined in the new  \textsc{Pythia}~6.1 event generator
\cite{pythianew}. The current version is 6.156, consisting of
some 53,000 lines of Fortran 77 code. The code, update notes,
the older long manual and sample main programs may be found at 
the webpage\\
\hspace*{\fill}
\texttt{http://www.thep.lu.se/}$\sim$\texttt{torbjorn/Pythia.html}
\hspace*{\fill}\\
while a new complete long manual is in preparation. 

A general-purpose high-energy physics event generator, such as
\textsc{Pythia}, need to contain a simulation of several physics 
aspects. 
\begin{Enumerate}
\item Hard subprocesses, such as $e^+ e^- \to Z^0 h^0$, which 
define the physics of interest.
\item Resonance decays, such as $Z^0$ and $h^0$ above,
which attach closely to the hard process itself, and often need to 
be simulated with full angular correlations.
\item Final-state parton showers to add a realistic multi-jet 
and internal jet structure (the former alternatively possible 
with higher-order matrix elements). 
\item Initial-state showers, in $e^+ e^-$ colliders mainly 
photon emission but for resolved-$\gamma$ events also of the 
QCD type encountered in hadronic collisions.
\item Multiple parton--parton interactions, for hadrons and 
resolved photons.
\item Beam remnants, left behind when not the full CM energy
is used in the hard process, again for hadrons or resolved
photons.
\item Hadronization, whereby the coloured quarks and gluons,
colour-connected into strings, are transformed into colour
singlet hadrons.
\item Normal secondary decays of unstable hadrons and leptons.
\item Potentially: interconnection effects, such as colour
rearrangement and Bose--Einstein effects, that complicate the 
hadronization/decay process.
\item The forgotten or unexpected, on the principle that a chain 
is never stronger than its weakest link.
\end{Enumerate}
Several of these areas have been improved in recent years. 

\textsc{Pythia} contains well over 200 different hard subprocesses,
in areas such as hard and soft QCD physics, heavy-flavour 
production, $W / Z$ production, prompt-photon production,
photon-induced processes, standard model Higgs production,
both for a light and a heavy Higgs state, non-standard Higgs 
particle production, production of new gauge bosons,
Technicolor production, compositeness effects, left--right 
symmetric models, leptoquark production and, last but not least,
Supersymmetric particle production.

The area of $\gamma\gamma$ physics has been significantly
expanded the last year, with the objective to present a 
`complete' framework of 
$\gamma\gamma / \gamma^*\gamma / \gamma^*\gamma^*$
interactions spanning the whole range of photon virtualities
$Q^2$, with special emphasis on the transition region when
$Q^2 \sim m_{\rho}^2$ \cite{gagawork}. This is the topic of a 
separate talk in these proceedings~\cite{gagatalk}.

The SUSY scenario contains a rather complete set of $2 \to 2$
processes in the MSSM, with $R$-parity conservation. News relative
to \textsc{SPythia} include an improved simulation of 
sbottom production, also including contributions from $b$'s
in the incoming hadrons, scenarios with the gravitino as the 
lightest SUSY particle, several 3-body decays of SUSY particles,
and the possibility of processes with displaced decay vertices.

The Technicolor simulation has progressed away from the simple
picture of $\rho_{\mathrm{tc}}$ resonance production with 
subsequent decay, to a more sophisticated one of $2 \to 2$
processes with full interference between $\rho_{\mathrm{tc}}^0$, 
$\omega_{\mathrm{tc}}^0$, $Z^0$ and $\gamma^*$ in the neutral 
$s$-channel and between $\rho_{\mathrm{tc}}^{\pm}$ and $W^{\pm}$ in 
the charged ones. The 2-body final states contain appropriate
combinations of $\pi_{\mathrm{tc}}^{\pm}$, $\pi_{\mathrm{tc}}^0$,
${\pi'}_{\mathrm{tc}}^0$, $W_{\mathrm{L}}^{\pm}$, 
$Z_{\mathrm{L}}^0$, $W^{\pm}$, $Z^0$ and $\gamma$ particles
(where index L on $W^{\pm}$ and $Z^0$ denotes longitudinal states).

The ${Z'}^0$ can now have flavour-dependent couplings different
between the first, second and third generations.

Higgs pair production, such as $h^0 A^0$ and $h^0 H^{\pm}$, is
now included as explicit processes; previously only some of them
could be accessed via ${Z'}^0/Z^0/\gamma^*$ production.

In the context of a left--right symmetric scenario, i.e. with an 
extra ${\mathbf{SU(2)}}_R$ symmetry group, doubly-charged Higgs
states appear, one set for each of the two $\mathbf{SU(2)}$ groups. 
$H^{\pm\pm}$ production singly or in pairs is complemented by 
decays to like-sign leptons or $W_L$'s or $W_R$'s, where the latter 
can decay further either to ordinary quarks or a lepton in association 
with a right-handed neutrino.

In compositeness scenarios, a few new processes have been added for
$e^*$ and $q^*$ production.

The area of extra dimensions is still fairly new, with many new 
scenarios appearing. Currently the program only contains a
narrow graviton resonance $G^*$ in the context of the
Randall--Sundrum model \cite{RS}. It is produced by $q \overline{q}$
or $g g$ fusion (with separate access to the high-$p_{\perp}$ tail  
by $q \overline{q} \to G^* g $, $q g \to G^* q$ and $g g \to G^* g$
in preparation), and can decay to a pair of fermions or gauge bosons.
 
Moving away from the topic of hard/primary processes, the major upgrade
of the generator has been in the area of final-state showers 
\cite{newshow}. This is further described in a separate talk in these
proceedings \cite{showtalk}. The starting point is the calculation 
of the matrix elements for gluon emission in two-body decays.   
Using the standard model and the minimal supersymmetric extension
thereof as templates, a wide selection of colour and spin structures
have been addressed, exemplified by $Z^0 \to q\overline{q}$,
$t \to b W^+$, $H^0 \to q\overline{q}$, $t \to b H^+$,
$Z^0 \to \tilde{q}\overline{\tilde{q}}$, $\tilde{q} \to \tilde{q}' W^+$, 
$H^0 \to \tilde{q}\overline{\tilde{q}}$, $\tilde{q} \to \tilde{q}' H^+$, 
$\tilde{\chi} \to q\overline{\tilde{q}}$, $\tilde{q} \to q\tilde{\chi}$,
$t \to \tilde{t}\tilde{\chi}$, $\tilde{g} \to q\overline{\tilde{q}}$, 
$\tilde{q} \to q\tilde{g}$, and $t \to \tilde{t}\tilde{g}$. 
The mass ratios $r_1 = m_b / m_a$ and $r_2 = m_c/m_a$ 
have been kept as free parameters. These matrix elements are then used 
to correct the gluon emission rate in showers. A modified choice of shower 
variables simplifies this operation. By applying the matrix element
corrections to all gluons emitted from the two primary decay products,
in suitably modified form, the process- and mass-dependent emission rate
should be well modelled also in the collinear region. With the modified
algorithm, a good description is obtained of mass effects in the gluon
emission rate at LEP1, i.e. of the difference between $b$-tagged and 
light quark jets. Predictions include a roughly 10\% higher three-jet
rate in Higgs decays to $b\overline{b}$ than for a $\gamma^*/Z^0$ of the
same mass and in the same decay channel. In top decays, the amount of 
radiation in the $W$ hemisphere is reduced relative to the older
algorithm.
  
In the area of hadronization, the Lund string model \cite{string} 
remains essentially unchanged. Improvements have been introduced in the 
low-mass region, however \cite{cluster}. Such strings can be produced by
parton-shower branchings $g \to q\overline{q}$ and, of course, in
$\gamma\gamma$ events. The low-mass strings can only produce one or two 
hadrons. The relative rate of these two possibilities, as a function of
string mass, has been modified.
When two hadrons are chosen to be produced, their relative orientation 
remains isotropic just at threshold, but is now matched to fragmentation
along the string direction further above threshold. If instead a collapse
occurs to a single hadron, the necessary shuffling of energy and momentum
is relative to other nearby string pieces.

A number of other changes and improvements have been made 
\cite{pythianew}. They include
\begin{Itemize}
\item many improved resonance decays,
\item newer parton distribution sets,
\item QED radiation also off an incoming muon,
\item an energy-dependent $p_{\perp\mathrm{min}}$ in multiple interactions,
\item the colour rearrangement options for $W^+W^-$ and $Z^0Z^0$ events
included in the standard code,
\item the Bose--Einstein algorithm has been expanded with new options,
\item a new baryon production scheme is allowed as an option,
\item simple routines for one-dimensional histogramming, and
\item standard interfaces to 2-, 4- and 6-fermion electroweak generators
and to 4-parton QCD generators, for performing subsequent showers and
hadronization. 
\end{Itemize}

In the near future, \textsc{Pythia}~6 will continue to develop, e.g. by
the inclusion of new processes and by further improvements of the showering
algorithms, and of course by bug fixes. In the longer future,
a radically new version of the program is required. Given the decisions
by the big laboratories and collaborations to discontinue \textsc{Fortran}
and instead adopt \textsc{C++}, it is natural to attempt to move also
event generators in that direction. User-friendly interfaces will have to
hide the considerable underlying complexity from the non-expert.
The \textsc{Pythia}~7 project got going three years ago, and is an effort 
to reformulate the event generation process in object oriented language. 
A strategy document \cite{leifone} was followed by a first `proof of 
concept version' in June 2000 \cite{leiftwo}, containing the generic
event generation machinery, some processes, and the string fragmentation
routines. In the next few years, the hope is to produce useful versions, 
even if still limited in scope. Due to the considerable complexity of the
undertaking, it will still be several years before the \textsc{C++} version
of \textsc{Pythia} will contain more and better physics than the
\textsc{Fortran} one. A corresponding project exists for a 
\textsc{Herwig++}, with two postdocs dedicated to the task, and some 
code sharing with that project is likely.

\end{document}